\documentclass[12pt]{article}

\usepackage{graphicx}
\usepackage{dcolumn}
\usepackage{bm}
\usepackage[            
pdftex,                 
pdfborder= 0 0 0,
a4paper,                
colorlinks=true,        
citecolor=green,        
linkcolor=cyan,         
menucolor=blue,         
urlcolor=magenta,       
bookmarksopen=true
]{hyperref}

\usepackage{amsmath}
\usepackage{amssymb}

\newcommand{\z}{Z\kern-0.45emZ}
\newcommand{\vi}{I\kern-0.3emB}
\newcommand{\1}{I\kern-0.3emI}

\newcommand{\e}{I\kern-0.3emE}
\newcommand{\E}{I\kern-0.3emE}
\newcommand{\ind}{I\kern-0.3emI}
\newcommand{\p}{I\kern-0.3emP}

\newcommand{\be}{\begin{equation}}
\newcommand{\ee}{\end{equation}}

\begin{document}


\title{The Taylor-Frank method cannot be applied to some biologically important, continuous fitness functions}

\author{Roberto H. Schonmann$^{1,2}$, Robert Boyd$^{3}$, Renato Vicente$^2$}

\date{\today}

\maketitle

\begin{center}
1. Dept. of Mathematics, University of California at Los Angeles,
CA 90095, USA \\
2. Dept. of Applied Mathematics, Instituto de Matem\'atica e Estat{\'\i}stica,
Universidade de S\~ao Paulo, 05508-090, S\~ao Paulo-SP, Brazil\\
3. Dept. of Anthropology, University of California at Los Angeles,
CA 90095, USA
\end{center}




\vspace{10mm}

\begin{abstract}
The Taylor-Frank method for making kin selection models when fitness is a nonlinear function of a continuous phenotype 
requires this function to be differentiable. This assumption sometimes fails for biologically important fitness functions, for instance in microbial data and the theory of repeated n-person games, even  when fitness functions are smooth and continuous. In these cases,  the Taylor-Frank methodology cannot be used, and a more general form of direct fitness must replace the standard one to account for kin selection, even under weak selection.
\end{abstract}

\vspace{10 mm}



\vspace{10 mm}

\noindent{\bf Acknowledgments:}
We are grateful to Clark Barrett for enlightening discussions.
We are also grateful to  Sam Bowles,
Steve Frank,
Anne Kandler,
Laurent Lehmann,
and Jeremy Van Cleve
for nice conversations and useful feedback on various aspects of this project and related subjects. This project was
Partially supported by CNPq, under grant 480476/2009-8.

\vspace{10mm}
\noindent{\bf Emails: }{\href{mailto:rhs@math.ucla.edu}{rhs@math.ucla.edu}, \
\href{mailto:rboyd@anthro.ucla.edu}{rboyd@anthro.ucla.edu}, \
\href{mailto:rvicente@ime.usp.br}{rvicente@ime.usp.br}




\newpage

\section{Introduction}
According to Hamilton's rule, the fitness of an allele should be measured by how much it affects the
reproductive success of its carriers, added to the effect that its carriers have on
the reproductive success of others weighted by relatedness \cite{Ham64}.  In its original formulation,
Hamilton's rule required additive fitness effects
\cite{Grafen85}, and a number of extensions have been developed to deal
with nonlinearities (reviewed, e.g., in \cite{Frank, WGF, GWW}).
In one of
the most influential extensions,
Taylor and Frank \cite{TF}, it is assumed that (1) fitnesses are functions of
continuously varying phenotypes, and (2) phenotypic variation is small enough that linear approximations to the nonlinear fitness functions are accurate. This approach has been usefully applied to a wide range of biological problems
(see, e.g., \cite{Frank}, \cite{LR}, \cite{WGF} Box 6.1 and
\cite{GWW} Box 6), and
it has been suggested
(see e.g., \cite{Frank} pp. 35-36 , \cite{WGF} pp. 137-138 and \cite{GWW} Box 6)
that it shows that Hamilton's rule,
in terms of marginal costs and benefits,
can always be applied to problems with continuously varying phenotypes,
as long as variation in the traits is small enough. 
However, as we explain in Section 2, the Taylor-Frank method depends on the assumption that fitness functions are
differentiable as functions of two variables: a focal individual's phenotype and the average phenotype in
its social environment.
And in Section 3 we point out that continuous fitness functions that occur in real biological applications may not be differentiable. When this is the case one cannot use the Taylor series approximation (and therefore the chain rule
of multi-variable calculus) that is the basis of the Taylor-Frank method.
These observations are particularly relevant because several important treatments of social evolution,
following on the steps of \cite{TF},
assume (see, e.g., p. 95 in \cite{Rousset}) that fitness functions are differentiable.
We will explain also in Section 2,
how the Taylor-Frank direct fitness method can be generalized so that
kin selection can be applied to problems for which one
cannot make this assumption.

\section{Invasion by a rare mutant under weak selection}
The Taylor-Frank direct fitness approach assumes that the fitness of an individual
is affected by its own phenotype and the phenotypes of other
individuals in its social environment. Individual phenotype is represented by a heritable quantitative
character $y$. For instance, $y$ could represent the amount of some costly to produce substance that
the individual secretes in the environment and that is beneficial to nearby individuals.
Initially all individuals in the population have the same value of this character, $y=\bar y$.
Rare mutations produce a variant with $y = \bar y + \delta$, where $|\delta|$ is small.
The rare mutant will invade if it has higher average fitness than the wild type.
To compute these average fitnesses, randomly select a focal individual from
the population, and let $z$ be the average phenotype in the social environment of this focal individual.
If the focal is a wild type, all the individuals in its social environment are likely to be wild types,
and therefore, $z = \bar y$. But if the focal individual is a mutant, other individuals in
its social environment may also be mutants, for instance
due to common descent. Let $z = yX + \bar y (1-X) = Z$, where $X$ is the
random variable that represents the fraction of members of the social environment that are
mutants. Denote by $w(y,z)$ the fitness of a focal individual with phenotype $y$, in a social
environment in which the average phenotype is $z$. The average fitness of the wild types is
then $w^{\mbox{w}} = w(\bar y, \bar y)$, while that
of the mutants is $w^{\mbox{m}} = \E w(y,yX + \bar y (1-X))$, where the expectation (denoted by $\E$) corresponds
to an average over social environments, i.e., over $X$. The mutants will
invade the population  when $w^{\mbox{m}} - w^{\mbox{w}} > 0$.
Because $|\delta|$ is small, one needs only to consider
what happens in the neighborhood of the point where $y = z = \bar y$. Key to the Taylor-Frank approach is the
assumption that the chain rule of multi-variable calculus
applies and gives, neglecting terms that are much smaller than $|\delta|$,
\be
w^{\mbox{m}} - w^{\mbox{w}} \ = \ \delta \, \E \left\{ \left.\frac{\partial w}{\partial y} \right|_{y= z = \bar y}
\, + \, X \, \left. \frac{\partial w}{\partial z} \right|_{y= z = \bar y} \right\}
\ = \ \delta (-C + B R),
\label{TF}
\ee
where $-C$ and $B$ are the values of the partial derivatives in the $y$ and $z$ directions at the
point $y=z=\bar y$ and $R = \E(X) = (\E(Z)-\bar{y})/(y-\bar{y})$
is the average relatedness in the social environment.
Provided that one can apply the chain rule, as above, the conclusion is that Hamilton's rule
\be
C < BR
\label{Hamilton}
\ee
is the necessary and sufficient condition for the mutants to invade the
monomorphic population with phenotype $\bar y$.

However, the use of the chain rule requires
(see, e.g., \cite{Kaplan})
that the function $w(y,z)$ be differentiable,  meaning that it is
well approximated by a linear function of $y$ and $z$, in
the neighborhood of $(\bar y, \bar y)$. That is, 
up to an error term that is much smaller than $|y-\bar y| + |z - \bar z|$,
we must have in good approximation $w(y,z) = \alpha + \beta y + \gamma z$, in the neighborhood
of $(\bar y, \bar y)$.
In other words, the surface that represents the function $w(y,z)$ must be well approximated
by a plane in the neighborhood of $(\bar y, \bar y)$.
In a one-dimensional setting, differentiability just means that the function is smooth and without 
kinks, and this is the same as being well approximated by a straight line, close to a given point.  In two or higher dimensions, functions can be smooth and free of kinks, but not approximated 
by a plane, and therefore not differentiable.  To intuitively understand
this well known idea, and see its biological meaning,
assume only that
for every value of $x$ in the range from 0 to 1,
the directional derivative of $w(y,z)$,
\be
v(x) \ = \ \left. \frac{d w(y, yx + \bar y (1-x))}{dy} \right|_{y = \bar y}
\label{v}
\ee
in the direction of the straight line
$z = yx + \bar y (1-x)$,
is well defined.
This directional derivative gives the incremental fitness effect
of changes in $y$
for a given fixed fraction $x$ of mutants in the social environment. These derivatives can exist for every value 
of $x$ and at the same time $w(y,z)$ is not differentiable $(y,z)$---see Fig(\ref{figTF}) for an example.
More formally, neglecting an error term much smaller than $\delta$, we have
$w(y,z) - w(\bar y, \bar y) = \delta v(x)$, when $y = \bar y + \delta$ is close to
$\bar y$ and $(z - \bar y)/(y - \bar y) = x$.
The quantity $v(x)$ is therefore the marginal fitness of a focal mutant, in a social environment
with a fraction $x$ of mutants.
Differentiability of the function of two variables $w(y,z)$ at
$(\bar y, \bar y)$ implies,
through the chain rule, that $v(x)$ must be the
linear function of $x$ given by $v(x) = -C + Bx$, with
$-C = \left. \partial w / \partial y \right|_{y=z=\bar{y}}$
and
$B = \left. \partial w / \partial z \right|_{y=z=\bar{y}}$.
This condition may or not hold in biologically relevant situations
(see Section 3).

This means that the Taylor-Frank approach
does not apply to situations in which the marginal fitness $v(x)$ of
the mutants is a non-linear function of the \emph{fraction} $x$ of mutant individuals in the social
environment.
There is nevertheless no difficulty in obtaining a valid (direct fitness, kin selection)
condition for invasion, based
only on the assumption that the directional derivatives $v(x)$ exist.
The quantity
$\delta v(x)$ is close to the difference in fitness between a mutant focal
individual in a social environment with a fraction $x$ of mutant individuals and the
fitness of a wild focal individual in an environment in which everyone is of wild type.
Therefore, when $\delta$ is small,
neglecting terms that are much smaller than $|\delta|$, we have
\be
w^{\mbox{m}} - w^{\mbox{w}} \ = \ \delta \, \E \left(v(X)\right).
\label{genH}
\ee
The
condition for the mutants to invade the
monomorphic population with phenotype $\bar y$ is therefore
\be
\E \left(v(X)\right) > 0,
\label{gHamilton}
\ee
which generalizes Hamilton's rule (\ref{Hamilton}), and reduces to it precisely when
$v(x) = -C + B x$ is a linear function of $x$, i.e., when the interaction of
the individuals in each environment affects fitnesses of mutants as a linear public goods game.
We show in Section 4 that
the same conclusion holds when costs and benefits are conceptualized as the coefficients in a regression of
fitness against phenotypic value.

The contrast between the simplicity and apparent generality in the derivation of (\ref{TF})
and the limitations explained in the previous paragraph may seem puzzling at first sight.
This apparent paradox is solved once one
understands that condition (\ref{TF}) relies on the assumption that $w(y,z)$ is differentiable
in the neighborhood of the point where $y=z=\bar y$, and that this means that the surface
that represents this function is well approximated by a plane close to that point.
To see why this assumption fails, consider Fig. \ref{figTF} in which the
difference in the fitness of mutants and wild types,
$w(y,z) - w(\bar y, \bar y) \approx \delta v(x)$,
is a sigmoidal function of the fraction of mutants, $x$, for each given
value of $\delta$. As the value of $\delta$ decreases, the values of
$|w(y,z) - w(\bar y, \bar y)|$
approach 0, but the sigmoidal shape does not change, implying that
also the limit $v(x)$
is sigmoidal, rather than a linear function.  This means that the surface representing $w(x,z)$
cannot be approximated by a plane, close to $(\bar y, \bar y)$.
Of course, as the usefulness of the
Taylor series approach throughout science attests, most nonlinear functions
of interest (in two variables) are well approximated by a
plane in the neighborhood of a point. When this happens in our setting,
the Taylor-Frank approach is correct. However,
as we explain in the next section, there are important biological applications for which data or theory
indicate that this is not the case and (\ref{TF}) and (\ref{Hamilton}) are
not a good approximation for (\ref{genH}) and (\ref{gHamilton}), which instead are
proper expressions of kin selection in those cases (assuming rarity of the mutant type and small
trait variability, i.e., small $|\delta|$).

\section{Biological significance}
There are at least two biological contexts in which fitness appears to be a nonlinear function of the fraction (or number)
of different types in the social environment.

First, experimental evidence from micro-organisms \cite{GYO, CRL, SDZ} indicates that
$v(x)$ sometimes is non-linear in $x$. However these data may be consistent with other functional forms.
One could argue that experimental data does not refer to a limit in which $\delta \to 0$, and that
the data comes from situations in which selection may be strong.
This raises the question of how small $\delta$ has to be for one to regard selection
as weak. Basically selection is weak when the differences in phenotype in the population
produce only minor differences in reproductive success, so that one can compute
(\ref{genH}) assuming that the expectation corresponds to neutral drift without selection.
(Separation of time scales; see, e.g., \cite{Rousset, RR, Lessard, SVC}.) Whether $\delta$ can be that small
while $v(x)$ is empirically non-linear is an important question to be investigated
experimentally.

Second, in repeated n-player games successful strategies make cooperation contingent on behaviors of others in the group.
To see how nondifferentiable fitness functions arise in such repeated games, consider the iterated n-person prisoner's dilemma
(or public goods game) \cite{Joshi, BR, SVC}.
Social interactions of this kind are likely important in all kinds of social vertebrates, and especially primates.  Chimpanzee patrolling and human food sharing may be examples.  Suppose that individuals interact repeatedly in groups of size $n$, and the extent of individual prosocial action is a continuous variable (e.g. the amount of food shared, or the level of risk taken on).  Let the value of this variable for individual $j$ be $y_j$ and the fitness effect of one period of interaction for individual $j$ be $-cy_j+\frac{b}{n-1}\sum_{i \ne j} y_i = -cy_j+
b z_j$, where the sum is over the other members of individual $j$'s group.
Individual behavior is contingent.  The wild type give $\bar y$ on the first interaction, and continue to give during the remaining $T$ interactions as long as a fraction $\theta$ of the other group members give at least $\bar y$.  However there is a rare invading type that gives $\bar y+\delta$, where $\delta$ is small, and continues to give this amount
as long as a fraction $\theta$ of the other group members give at least $\bar y+\delta$.
As before, if a focal individual is a mutant, it has in its social environment a fraction
$x = (z - \bar y)/(y - \bar y)$ of mutants.
The fitness function takes the form $w(y,z) = w_0 + (y - \bar y)(-c + bx)$, if $x < \theta$ and
$w(y,z) = w_0 + (y - \bar y)(-c + bx)T$, if $x \geq \theta$, where $w_0 = w(\bar y,\bar y)$ is a constant.
This yields the non-linear marginal fitness function
$v(x) = -c + b x$, if $x < \theta$, and $v(x) = (-c + b x)T$, if $x \geq \theta$.
No matter how small $\delta$ is,
the marginal effect of changes in $y$ and $z$ on the fitness of rare types depends on whether $x$ is greater than or less than $\theta$.
Contingent behavior that leads to non-linear fitness functions is a common feature in the modeling
of social evolution, especially of human cooperation; see, e.g., \cite{BG} and references therein.

 Hamilton's rule (\ref{Hamilton}) is appealing because the only information needed about patterns of interaction
is the relatedness $R$. Assuming $\delta$ small enough, $R$ can be obtained from the distribution of neutral genetic markers in the same population. That is, there is a separation of time
scales so that changes due to demographic processes occur much faster than changes due to selection. When (\ref{gHamilton}) has to replace (\ref{Hamilton}), $R$ is not enough.  More detailed information is needed about the distribution of $X$.
However as long as selection is weak, the separation of time scales  exists and the distribution of $X$ can be calculated
 using distribution of neutral genetic markers.  Problems of this type have been
addressed in a number of papers, including \cite{RR, Lessard, Ohtsuki, SVC}.
This approach was applied in
\cite{SVC} to the iterated
public goods game, in a population structure for which it was shown that 
the distribution of $X$ is a beta distribution,
with parameters specified by
the level of gene flow (group size $\times$ migration rate).
Under biologically plausible assumptions, the generalization of Hamilton's
rule given above (\ref{gHamilton}) yielded the invasion condition
$-\ln(1 - c/b) < R \ln T$, when $T$ is large, illustrating the usefulness of (\ref{gHamilton}).
While certainly more complicated than (\ref{Hamilton}), it can be analyzed
in detail in some important cases, and provides transparent conditions for invasion
of a rare mutant.

\section{Regression coefficients}
The invasion condition $w^{\mbox{m}} - w^{\mbox{w}} > 0$
can be expressed in terms
of regression coefficients. Define $y_{\bullet}$, $z_{\bullet}$ and
$w_{\bullet} = w(y_{\bullet}, z_{\bullet})$ as the random variables that are equal to the values
that these quantities take for the focal individual. The invasion condition is then 
(see, e.g., \cite{GWW} display (5), or \cite{WGF} display (6.5))
\be
\beta_{w_{\bullet},y_{\bullet}|z_{\bullet}} \, + \, \beta_{z_{\bullet},y_{\bullet}} \,
\beta_{w_{\bullet},z_{\bullet}|y_{\bullet}} \ > \ 0.
\label{reg2}
\ee
Where the regression coefficients are defined as the numbers that together with the proper choice of
the constants $\alpha'$ and $\alpha''$ minimize
\be
\E \left( \left(\alpha' + \beta_{z_{\bullet}, y_{\bullet}} \, y_{\bullet} - z_{\bullet}\right )^2 \right)
\ \ \ \ \ \mbox{and} \ \ \ \ \
\E \left( \left(\alpha'' + \beta_{w_{\bullet},y_{\bullet}|z_{\bullet}} \, y_{\bullet}
+ \beta_{w_{\bullet},z_{\bullet}|y_{\bullet}} \, z_{\bullet}
- w_{\bullet}\right )^2 \right)
\label{minsquare2}
\ee
This definition in (\ref{minsquare2}) says that $\beta_{w_{\bullet},y_{\bullet}|z_{\bullet}}$, $\beta_{w_{\bullet},z_{\bullet}|y_{\bullet}}$  and $\alpha''$ are the numbers that
make the  function
$f(y,z) \, = \, \alpha'' \,  + \, \beta_{w_{\bullet},y_{\bullet}|z_{\bullet}} \, y \, + \,
\beta_{w_{\bullet},z_{\bullet}|y_{\bullet}} \, z$ the best linear approximation to the function $w(y,z)$
(in the sense that it minimizes the square of errors weighted by probabilities
over the values of $y$ and $z$). The condition (\ref{reg2}) is appealing because
$\beta_{z_{\bullet},y_{\bullet}} = R$ is the relatedness in the social environments, and therefore
(\ref{reg2}) is equivalent to
\be
\beta_{w_{\bullet},y_{\bullet}|z_{\bullet}} \, + \, R \,
\beta_{w_{\bullet},z_{\bullet}|y_{\bullet}} \ > \ 0.
\label{reg3}
\ee

If $w(y,z)$ is differentiable, and distribution of values of $(y_{\bullet}, z_{\bullet})$ is narrowly concentrated close
to $(\bar{y},\bar{y})$, then the regression coefficients in (\ref{reg3}) are the same as the marginal fitnesses
 derived using the Taylor-Frank method. If $w(y,z)$ is differentiable at $(\bar{y},\bar{y})$, $w(y,z)$ is well approximated by a linear function of $y$ and $z$, in the neighborhood of this point.
This means that
\be
w(y,z) = A -Cy +Bz + o(|y-\bar{y}| + |z-\bar{y}|),
\label{diff}
\ee
with
$-C = \left. \partial w / \partial y \right|_{y=z=\bar{y}}$
and
$B = \left. \partial w / \partial z \right|_{y=z=\bar{y}}$.
Thus $w_{\bullet} = A -C y_{\bullet} + Bz_{\bullet}$ is a good approximation
and hence the second optimization problem in (\ref{minsquare2}) is solved  by
$\beta_{w_{\bullet},y_{\bullet}|z_{\bullet}} = -C$ and
$\beta_{w_{\bullet},z_{\bullet}|y_{\bullet}} = B$,
regardless of the details of the
joint distribution of $y_{\bullet}$ and $z_{\bullet}$. This approximation becomes better and
better, as $\delta \to 0$, and therefore (\ref{reg2}) is well approximated by Hamilton's condition $C < BR$,
in the limit of weak selection.

But suppose now that $w(y,z)$ is not differentiable at $(\bar{y},\bar{y})$.
In this case, we can even add
the assumption that the mutant types, with $y = \bar{y} + \delta$, are rare, and still we
will not have the approximate equalities between the regression coefficients
$\beta_{w_{\bullet},y_{\bullet}|z_{\bullet}}$
and
$\beta_{w_{\bullet},z_{\bullet}|y_{\bullet}}$
and, respectively, the partial derivatives
$\left. \partial w / \partial y \right|_{y=z=\bar{y}}$
and
$\left. \partial w / \partial z \right|_{y=z=\bar{y}}$.
To illustrate this point, suppose that $w(y,z)$ is given in Fig 1. The assumptions that we made about
$\delta$ being small and the mutants being rare, implies that the distribution of
$(y_{\bullet}, z_{\bullet})$ concentrates close to the point $(\bar{y},\bar{y})$.  But the
distribution over this segment depends on demographics---it is determined by the distribution of the random
variable $X$ that gives the number of mutants in the social environment of a mutant focal.
Because $w(y,z)$ is not well approximated by a linear function of $y$ and $z$ in the relevant region,
(even for very small values of $\delta$), the regression coefficients
$\beta_{w_{\bullet},y_{\bullet}|z_{\bullet}}$
and
$\beta_{w_{\bullet},z_{\bullet}|y_{\bullet}}$
will depend on the distribution of $X$ in a substantial way.
To see why consider the function, $v(x)$,  shown in  Fig 1.  This function
is very flat when $x$ is close to 0 or 1, but is steeply increasing when $x$ takes
intermediate values. Now, compare three scenarios.
(1)
If the distribution of $X$ is concentrated close to $x=0$, then we will have
$\beta_{w_{\bullet},y_{\bullet}|z_{\bullet}}$ close to
$\left. \partial w / \partial y \right|_{y=z=\bar{y}}$,
which is a negative number,
and
$\beta_{w_{\bullet},y_{\bullet}|z_{\bullet}}$ close to 0.
(2) If the distribution of $X$ is concentrated close to $x=1$, then we will have
$\beta_{w_{\bullet},y_{\bullet}|z_{\bullet}}$ positive and again
$\beta_{w_{\bullet},z_{\bullet}|y_{\bullet}}$ close to zero.
(3) If the distribution of $X$ is concentrated in intermediate values of $x$, then
we will have
$\beta_{w_{\bullet},y_{\bullet}|z_{\bullet}}$ even more negative than
$\left. \partial w / \partial y \right|_{y=z=\bar{y}}$, and
$\beta_{w_{\bullet},z_{\bullet}|y_{\bullet}}$ large and positive.

The idea that when selection is weak and mutants are rare (vanishing trait variation
in the population) we would have in good approximation
$\beta_{w_{\bullet},y_{\bullet}|z_{\bullet}}
= \left. \partial w / \partial y \right|_{y=z=\bar{y}}$
and
$ \beta_{w_{\bullet},z_{\bullet}|y_{\bullet}}
= \left. \partial w / \partial z \right|_{y=z=\bar{y}}$
has been claimed often (e.g., \cite{GWW} Box 6 and \cite{WGF}
as they justify deriving (6.7) from (6.5)).
This idea is intuitive and appealing, but unfortunately it is
not correct, unless $w(y,z)$ is differentiable in the relevant region.

\section{Conclusions}
Whether the Taylor-Frank method is appropriate depends on the biological facts describing how the fitness of a mutant individual, with a small mutation, depends on the fraction $x$ of individuals in its social environment that carry the same
mutation. The method can be properly applied only when this dependence is linear.
However, even when the Taylor-Frank method is not appropriate, kin selection under weak selection and
rarity of the invading mutant is properly
described by the more general (\ref{genH}), and the corresponding generalized Hamilton
rule (\ref{gHamilton}).

\begin{figure}[th!]
\begin{center}
\includegraphics[width=15cm]{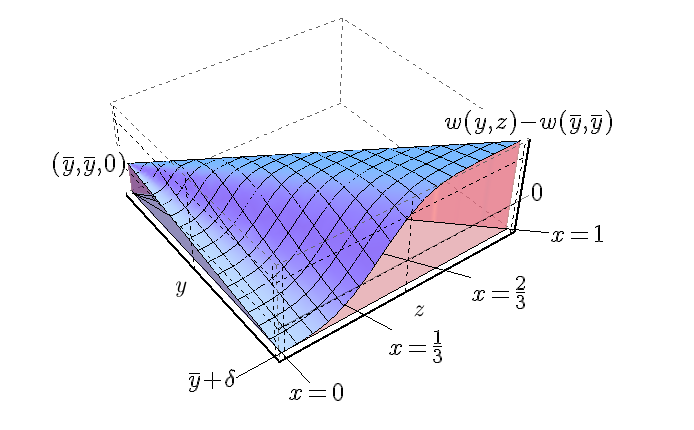}
\end{center}
\caption{The surface representing $w(y,z) - w(\bar y, \bar y)$
in the neighborhood of the point $(y,z) = (\bar y, \bar y)$. This point appears in the
left side of the picture, and the function takes the value 0 there. In the picture $y$ ranges from $\bar y$
to $\bar y + \delta$, and $z$ ranges from $\bar y$ to $y$. The parameter $x = (z-\bar y)/(y -\bar y)$ identifies
directions in the $(y,z)$ plane, away from the point $(y,z) = (\bar y, \bar y)$,
and biologically represents the fraction of individuals in the 
social environment of a mutant focal individual that are also mutants.
The values of the
directional derivatives $v(x)$, which represent marginal fitnesses,
are indicated by the s-shaped curve
produced by the intersection of the surface with the plane $y = \bar y + \delta$
(this s-shaped curve appears as the frontal border of the blue surface in the
picture).
The surface would only be well approximated by a plane, in the neighborhood of
$(y,z) = (\bar y, \bar y)$,
if $v(x)$ were a linear function, rather than s-shaped.
Notice that $w$ is
not differentiable anytime that $v(x)$ is a non-linear function of the fraction of mutant-types in the
social environment; no kinks or discontinuities are necessary.
When, as in this picture, differentiability at $(y,z) = (\bar y, \bar y)$ fails,
one can not use the chain rule as in the derivation of (\ref{TF}), but the
more general (\ref{genH}) still applies and provides the direction of selection.
}
\label{figTF}
\end{figure}

\end{document}